\documentclass[submission,copyright,creativecommons]{eptcs}
\providecommand{\event}{ACL2 2022} 
\usepackage{breakurl}             
\usepackage{hyperref}
\usepackage{underscore}           
\usepackage[nocompress]{cite}

\usepackage{amsmath}
\usepackage{amssymb}
\usepackage{amsbsy}
\usepackage{hhline}
\usepackage{xcolor,colortbl}
\usepackage{mathpartir}
\usepackage{tikz}
\usetikzlibrary{arrows,calc,positioning,backgrounds,fit,shapes,shadows,trees}

\newcommand*{\connectorH}[4][]{
  \draw[#1] (#3) -| ($(#3) !#2! (#4)$) |- (#4);
}

\tikzset{
  basic/.style  = {draw,rectangle,font=\sffamily\fontsize{12}{12}\selectfont}, 
  root/.style   = {basic, font=\sffamily\fontsize{12}{12}\selectfont, rounded corners=2pt, thin, align=center,
                   fill=orange!30},
  level 0/.style = {basic, font=\sffamily\fontsize{12}{12}\selectfont, thin, align=center, fill=blue!10},
  level 1/.style = {basic, trapezium, trapezium left angle=70, trapezium right angle=110, font=\sffamily\fontsize{12}{12}\selectfont, thin, align=center, fill=green!20},
  level 2/.style = {basic, font=\sffamily\fontsize{12}{12}\selectfont, rounded corners=2pt, thin, align=left,
    fill=yellow!30},
  level 3/.style = {basic, font=\sffamily\fontsize{12}{12}\selectfont, rounded corners=2pt, thin, align=left,
    fill=blue!20, text width=12em},
  level 4/.style = {basic, font=\sffamily\fontsize{12}{12}\selectfont, thin, align=center, 
    fill=green!30},
  level 5/.style = {basic, font=\sffamily\fontsize{12}{12}\selectfont, thin, align=center, 
    fill=orange!30},
  level 6/.style = {basic, rectangle, font=\sffamily\fontsize{12}{12}\selectfont, thin, align=center, 
    fill=cyan!20},
   boxaround/.style={draw=violet, font=\sffamily\fontsize{12}{12}\selectfont, thick, dashdotted,
     inner sep=0.8em},
 >=latex
}

\usepackage{paralist}
\usepackage{boxedminipage} 
\usepackage{booktabs} 

\newcommand{\etc}{\textit{etc}}
\newcommand{\eg}{\textit{e.g.}}

\usepackage{scalerel}

\title{Hardware/Software Co-Assurance using the Rust Programming
  Language and ACL2}

\author{David Hardin
\institute{Collins Aerospace\\
Cedar Rapids, IA USA}
\email{david.hardin@collins.com}}

\def\titlerunning{Hardware/Software Co-Assurance using Rust}
\def\authorrunning{D.S. Hardin}

\begin{document}

\maketitle

\begin{abstract}

The Rust programming language has garnered significant interest and
use as a modern, type-safe, memory-safe, and potentially formally
analyzable programming language.  Our interest in Rust stems from its
potential as a hardware/software co-assurance language, with
application to critical systems such as autonomous vehicles.  We report
on the first known use of Rust as a High-Level Synthesis (HLS)
language.  Most incumbent HLS languages are a subset of C.  A
Rust-based HLS brings a single modern, type-safe, and memory-safe
expression language for both hardware and software realizations with
high assurance.  As a study of the suitability of Rust as an HLS, we
have crafted a Rust subset, inspired by Russinoff's Restricted
Algorithmic C (RAC), which we have imaginatively named Restricted
Algorithmic Rust, or RAR.  In our first implementation of a
RAR toolchain, we simply transpile the RAR source into RAC.  By so
doing, we leverage a number of existing hardware/software co-assurance
tools with a minimum investment of time and effort.  In this paper, we
describe the RAR Rust subset, detail our prototype RAR toolchain, and
describe the implementation and verification of several representative
algorithms and data structures written in RAR, with proofs of correctness
conducted using the ACL2 theorem prover.

\end{abstract}

\section{Introduction}

The Rust programming language has garnered significant interest and
use as a modern, type-safe, memory-safe, and potentially formally
analyzable programming language.  Google \cite{RustAndroid} and
Amazon \cite{SustainableRust} are major Rust adopters, and Linus
Torvalds has commented positively on the near-term ability of the
Rust toolchain to be used in Linux kernel development \cite{RustAndroidArs}.
The latter capability comes none too soon, as use of C/C++ continues
to spawn a seemingly never-ending parade of security vulnerabilities,
which continue to manifest at a high rate \cite{MSBugs} despite the
emergence and use of sophisticated C/C++ analysis tools.  Moreover,
the extremely aggressive optimizations in modern C/C++ compilers
have lead some researchers to declare that C is no longer suitable for
system-level programming \cite{CUnusable}, arguably C's major
\emph{rasion d'etre}.

Our interest in Rust stems from its potential as a hardware/software
co-assurance language.  This interest is motivated in part by emerging
application areas, such as autonomous and semi-autonomous
platforms for land, sea, air, and space, that require sophisticated
algorithms and data structures, are subject to stringent
accreditation/certification, and encourage hardware/software co-design
approaches. (For an unmanned aerial vehicle use case illustrating a formal
methods-based systems engineering environment, please consult 
\cite{case-models-2021}.)  In this paper, we explore the use
of Rust as a High-Level Synthesis (HLS) language \cite{HLS}.  Most incumbent HLS
languages are a subset of C, e.g. Mentor Graphics' Algorithmic C
\cite{AlgoC}, or Vivado HLS by Xilinx \cite{VivadoHLS}.  A Rust-based
HLS would bring a single modern, type-safe, and memory-safe
expression language for both hardware and software realizations,
with very high assurance.

As formal methods researchers, another keen interest is in being able
to reason about application-level logic written in the imperative
style favored by industry.  Much progress has been made to this end
in recent years, to the point that developers can verify the
correctness of common algorithm and data structure code that utilizes
common idioms such as records, loops, modular integers, and the like,
and verified compilers can guarantee that such code is compiled
correctly to binary \cite{cakeml:popl14}.  Particular progress has 
been made in the area of hardware/software co-design algorithms, 
where array-backed data structures are
common \cite{hardin-co-assurance, hardin-rac}.
(NB: This style of programming also addresses one of the shortcomings 
of Rust, namely its lack of support for cyclic data structures.)

As a study of the suitability of Rust as an HLS, we have crafted a
Rust subset, inspired by Russinoff's Restricted Algorithmic C
(RAC) \cite{Russinoff2022}, which we have imaginatively named
Restricted Algorithmic Rust, or RAR.  In fact, in our first
implementation of a RAR toolchain, we merely ``transpile'' 
(perform a source-to-source translation of) the
RAR source into RAC.  By so doing, we leverage a number of existing
hardware/software co-assurance tools with a minimum
investment of time and effort.  By transpiling  RAR to RAC, we gain
access to existing HLS compilers (with the help of some simple C
preprocessor directives, we are able to generate code for either the
Algorithmic C or Vivado HLS toolchains).  But most importantly for
our research, we leverage the RAC-to-ACL2 translator that Russinoff
and colleagues at Arm have successfully utilized in
industrial-strength floating point hardware verification.

We have implemented several representative algorithms and data
structures in RAR, including:

\begin{itemize}
\item{a suite of array-backed algebraic data types, previously
    implemented in RAC (as reported in \cite{hardin-rac});}
\item{a significant subset of the Monocypher \cite{monocypher} modern cryptography
    suite, including XChacha20 and Poly1305 (RFC 8439) encryption/decryption,
    Blake2b hashing, and X25519 public key cryptography; and}
\item{a DFA-based JSON lexer, coupled with an LL(1) JSON parser.  
   The JSON parser has also been implemented using Greibach
   Normal Form (previously implemented in RAC, as described in
   \cite{formal-filter-synth-langsec}).}
\end{itemize}

The RAR examples created to date are similar to their RAC counterparts
in terms of expressiveness, and we deem the RAR versions somewhat
superior in terms of readability (granted, this is a very subjective evaluation).

In this paper, we will describe the development and formal verification 
of an array-based set data structure in RAR.  Along the way, we will introduce
the RAR subset of Rust, the RAR toolchain, the array-based set example,
and detail the ACL2-based verification techniques, as well as the ACL2
books that we brought to bear on this example.  It is hoped that this
explication will convince the reader of the practicality of RAR as a
high-assurance hardware/software co-design language, as well as the
feasibility of the performing full functional correctness proofs of
RAR code.  We will then conclude with related and future work.

\section{An Aspirational Integrated Toolchain}

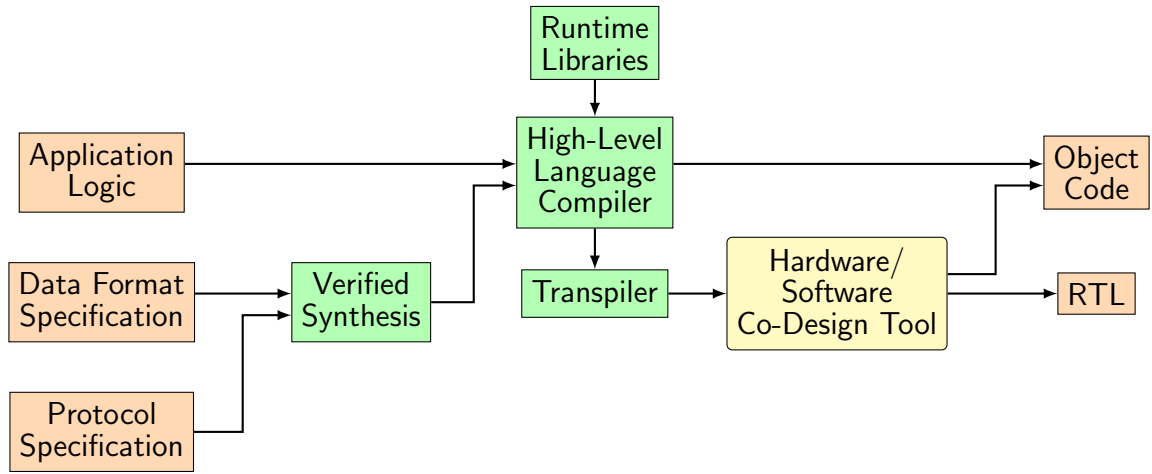
\begin{figure}
\center{
\begin{tikzpicture}[auto, scale=1.15]
\node[level 5] (AppLogic) at (0,3) [shape=rectangle,draw,align=center]{Application\\Logic};
\node[level 5] (DataSpec) at (0,1.5) [shape=rectangle,draw,align=center]{Data Format\\Specification};
\node[level 5] (ProtoSpec) at (0,0) [shape=rectangle,draw,align=center]{Protocol\\Specification};
\node[level 4] (VerSynth) at (3,1.5)[shape=rectangle,draw,align=center]{Verified\\Synthesis};
\node[level 4] (Transpiler) at (5.7,1.6)[shape=rectangle,draw,align=center]{Transpiler};
\node[level 4] (VerComp) at (5.7,3) [shape=rectangle,draw,align=center]{High-Level\\Language\\Compiler};
\node[level 4] (VerLib) at (5.7,4.5)[shape=rectangle,draw,align=center]{Runtime\\Libraries};
\node[level 2] (CoDes) at
(8.5,1.6)[shape=rectangle,draw,align=center]{Hardware/\\Software\\Co-Design Tool};
\node[level 5] (Obj) at (11.5,3) [shape=rectangle,draw,align=center]{Object\\Code};
\node[level 5] (RTL) at (11.5,1.6) [shape=rectangle,draw,align=center]{RTL};
\draw [->,thick] ([yshift=0.1cm]AppLogic.east) to ([yshift=0.1cm]VerComp.west);
\draw [->,thick] ([yshift=0.1cm]DataSpec.east) to ([yshift=0.1cm]VerSynth.west);
\connectorH [->,thick] {0.5} {ProtoSpec.east} {[yshift=-0.15cm]VerSynth.west};
\connectorH [->,thick] {0.5} {VerSynth.east} {[yshift=-0.15cm]VerComp.west};
\draw [->,thick] ([yshift=0.1cm]VerComp.east) to ([yshift=0.1cm]Obj.west);
\draw [->,thick] (Transpiler) to (CoDes);
\draw [->,thick] (CoDes) to (RTL);
\connectorH [->,thick] {0.5} {[yshift=0.225cm]CoDes.east} {[yshift=-0.15cm]Obj.west};
\draw [->,thick] (VerLib) to (VerComp);
\draw [->,thick] (VerComp) to (Transpiler);
\end{tikzpicture}}
\caption{Aspirational integrated hardware/software co-design, co-assurance toolchain.}
\label{fig:aspirational-toolchain}
\end{figure}

In order to place our research goals in context, let us consider an
aspirational integrated hardware/software co-assurance toolchain,
as shown in Fig.~\ref{fig:aspirational-toolchain}.

In this approach, developers can input various familiar high-level
specifications, and produce verified implementations for
those specifications, in software, hardware, or a combination of the
two.  Let us now consider each element of
Fig.~\ref{fig:aspirational-toolchain} in turn, proceeding left-to-right
and top-to-bottom.

\subsection{Inputs}

Many different common input specification types are anticipated, from
AADL models to algorithms in type-safe programming languages to ABNF
protocol specifications to lexer/parser specifications; only a few
representative input types are shown in
Fig.~\ref{fig:aspirational-toolchain}.  The input specification formats
for Verified Synthesis will be discussed further in
Section~\ref{sec:VerSynth}.  As for Application Logic, our approach
anticipates inputs in modern, type-safe language(s) supported by 
verified compilation, as described in Section~\ref{sec:VerComp}.

These input forms may be subject to certain subsetting rules, as not all
input specification utterances are appropriate for formal synthesis/analysis.
Additionally, input specifications may be augmented with
pragmas (often introduced as structured comments) to help guide
the execution of the formal tools described in the sections that follow.

\subsection{Verified Synthesis}
\label{sec:VerSynth}

Verified Synthesis constitutes a number of different verified
program synthesis tools for declarative specification input
forms commonly utilized by engineers and computer scientists,
including state machine specifications, protocol specifications (\eg,
ABNF), regular expressions, grammars, and the like.  Our goal is for
the Verified Synthesis tools to generate source code in the
language(s) of the High-Level Language Compiler(s) of
Section~\ref{sec:VerComp}.

\subsection{Runtime Libraries}

Runtime Libraries constitute a set of runtime services that a
verified application may require.  These include the usual runtime
services, as well as a number of common algebraic data types.  
Our goal is for the runtime libraries to be formally verified,
inspired by the verified runtime libraries for CakeML 
\cite{cakeml:popl14}.

Note that high-assurance design rules in particular domains may
require specializations of common runtime library services.  For
example, cyber-physical system designers generally limit the space
and time allocations for any given function, and require that algorithms
deliver results within a finite time, or suffer a watchdog timeout.
Furthermore, domain-specific high-assurance design rules,
such as mandated by RTCA DO-178C Level A \cite{DO-178C} for
flight-critical systems, frown on dynamic memory allocation,
preferring simple array-based data structure implementations.  This
discipline also benefits hardware/software co-design, as array-based
implementations are much easier to realize in hardware than dynamic
data structures, with their requirements for \emph{malloc} and
\emph{free} operations --- not to mention the attendant programming
errors that can result from dynamic memory management,
eg, use-after-free.  Thus, an algebraic data type may include a
high-level, functional implementation for which correctness proofs are
easier to obtain, as well as a fixed-memory version, accompanied by a
proof stating that the latter specializes the former.  These
specialization proofs could possibly be performed with the aid of
machine-learning-based automated proof refinement capabilities in
the future.

\subsection{High-Level Language Compiler}
\label{sec:VerComp}

We strongly advocate type-safe and memory safe modern languages,
preferably with verified compilers.  A few successful verified compilers have
been created in recent years, notably CompCert C \cite{leroy:cacm} and
CakeML \cite{cakeml:popl14}.  However, C is hardly an appealing
language for future high-assurance system development; a verified
compiler does little to address the many shortcomings of C from an
assurance perspective, including unrestricted pointer arithmetic, a
seemingly never-ending parade of buffer overflow and other memory
integrity/memory management vulnerabilities, unrestricted function
pointers, \etc.  On the other hand, Standard ML has a rather small
developer community, and its functional orientation makes it not
particularly well-suited for embedded systems programming.  The
F* dialect of ML, however, is the basis for some very interesting work
on high-assurance network protocols at Microsoft Research
\cite{EverParse} that shares many of our protocol verification objectives.

We aspire to provide verified compilation support for high-assurance
subsets of popular, modern functional/imperative hybrid languages, \eg,
Scala, Rust, or Swift.  These languages exhibit type safety, restrict
pointer operations, and are capable of producing efficient,
product-quality code.  An exemplary approach would include the 
creation of a \emph{verified compiler toolkit}, which would be used to
develop a set of verified compilers for semantically similar, but perhaps
syntactically quite different, languages.  Additionally, it is hoped
that the verified synthesis tools of Section~\ref{sec:VerSynth} can be
employed to build verified compilers using this toolkit approach.

\subsection{Transpiler}

A Transpiler in the context of Fig.~\ref{fig:aspirational-toolchain} is
a verified source-to-source translator that takes the development
source language(s) to a language that can be processed by the
Hardware/Software Co-Design Tool.  A Transpiler in our context
translates a higher-level language, \eg, Rust or Scala, to a
lower-level one, \eg, Algorithmic C; an unverified transpiler of this
sort has been created  by Robby of Kansas State University, and is
part of the DARPA CASE tools \cite{slang}.  It is hoped that the verified compiler
toolkit described in  Section~\ref{sec:VerComp} can be used to build
any needed transpilers with high assurance.

\subsection{Hardware/Software Co-Design Tool}

A Hardware/Software Co-Design tool, as the name implies, allows a
developer to allocate her designs to hardware, software, or a
combination of the two, producing software code and RTL as output.
Sophisticated co-design environments also allow the user to perform
``what if'' analyses, with the aid of simulation capability, adjusting
the allocations to hardware vs. software in order to optimize for
desired properties.  Example tools of this sort include Simulink
\cite{Simulink}, as well as the development environments for
``system'' languages such as SystemC, Algorithmic C, \etc.  The
Hardware/Software Co-Design Tool is not likely to be a verified
program itself (at least not initially), but should at least provide
the capability to export a design model that can be analyzed using
formal verification tools.

We have been evaluating a particular hardware/software design approach
employed by floating-point hardware verification researchers, detailed
further in Section~\ref{sec:algoc}.

\subsection{Object Code}

``Object Code'' is a broad term for the output of
compilation/assembly, from machine-independent virtual
machines (\eg, JVM \cite{JVM}, LLVM \cite{LLVM}) to the machine
code for a given  CPU (\eg, ARM, x86, RISC-V, PowerPC).  An object
code file is typically encoded in a binary format.  We are
specifically interested in compilation toolchains, such as CakeML,
that provide verified compilation down to the object code level; as
well as verified ``lifters'' that can take object code, and abstract
it to functions in a given logic, \eg, Myreen's Decompilation into
Logic work \cite{myreen:phd}.

\subsection{RTL (Register Transfer Logic)}

Register Transfer Logic (RTL) describes a hardware design at a
functional level, but at a sufficient level of fidelity that allows
for automated synthesis of a gate-level netlist, as well as
automated analysis and simulation.  This netlist can then 
be placed and routed to produce FPGA or ASIC implementations.
RTL is expressed using Hardware Description Languages (HDLs), such as
VHDL, Verilog, SystemVerilog, or Bluespec.

\section{RAC: Hardware/Software Co-Assurance at Scale}
\label{sec:algoc}

In order to begin to realize our aspirational vision for
hardware/software co-assurance at scale, we have 
conducted several experiments employing a state-of-the-art
toolchain, due to Russinoff and O'Leary, and originally designed for use 
in floating-point hardware verification \cite{Russinoff2022}, 
to determine its suitability for the creation of
safety-critical/security-critical applications in various domains.
Note that this toolchain has already demonstrated the capability to
scale to industrial designs in the floating-point hardware design and
verification domain, as it has been used in design verifications for
CPU products at both Intel and Arm.

Algorithmic C \cite{AlgoC} is a High-Level Synthesis (HLS) language,
and is supported by hardware/software co-design environments from
Mentor Graphics, \eg, Catapult \cite{Catapult}.  Algorithmic C defines
C++ header files that enable compilation to both hardware and software
platforms, including support for the peculiar bit widths employed, for
example, in floating-point hardware design.

\begin{figure}
\center{
\begin{tikzpicture}[auto, scale=1.2]
\node[level 5] (Hdr) at (2,3.2) [shape=rectangle,draw,align=center]{Algorithmic C\\Headers};
\node[level 5] (AlgoC) at (2,1.7) [shape=rectangle,draw,align=center]
{Algorithmic  C\\Source};
\node[level 0] (C++) at (1,0) [shape=rectangle,draw,align=center]{C++\\Compiler};
\node[level 0] (HW) at (3,0) [shape=rectangle,draw,align=center]{Hardware\\Synthesis};
\node[level 2] (Xlate) at (4.7,1.7) [shape=rectangle,draw,align=center]{ACL2\\Translator};
\node[level 4] (Lemmas) at (7,3.2) [shape=rectangle,draw,align=center]{Lemmas};
\node[level 2] (ACL2) at (7,1.7) [shape=rectangle,draw,align=center]{ACL2\\Theorem\\Prover};
\node[level 4] (Proofs) at (7,0.1) [shape=rectangle,draw,align=center]{Proofs};
\draw [->,thick] (Hdr) to (AlgoC);
\draw [->,thick] (AlgoC) to (C++);
\draw [->,thick] (AlgoC) to (HW);
\draw [->,thick] (AlgoC) to (Xlate);
\draw [->,thick] (Xlate) to (ACL2);
\draw [->,thick] (Lemmas) to (ACL2);
\draw [->,thick] (ACL2) to (Proofs);
\end{tikzpicture}}
\caption{Restricted Algorithmic C (RAC) toolchain.}
\label{RAC:toolchain}
\end{figure}
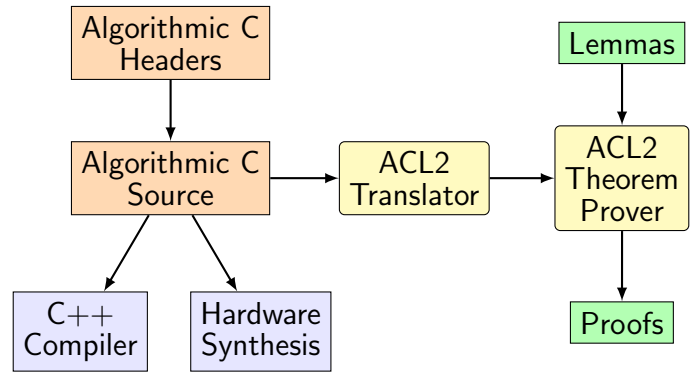

The Russinoff-O'Leary Restricted Algorithmic C (RAC)
toolchain, depicted in Fig.~\ref{RAC:toolchain}, translates a subset of
Algorithmic C source to the Common Lisp subset supported by the 
ACL2 theorem prover, as augmented by Russinoff's Register Transfer
Logic (RTL) books.

\begin{table}
  \center{
  \begin{normalsize}
\begin{tabular}{|c|c|}
  \hline
  \emph{Formal Verification ``Comfort Zone''} & \emph{Real-World Development} \\ \hhline{|=|=|}
  Functional programming & Imperative programming \\ \hline
  Total, terminating functions	& Partial, potentially non-terminating functions \\ \hline
  Non-tail-recursive functions & Loops \\ \hline
  Okasaki-style pure functional algebraic data types & Structs, Arrays \\ \hline
  Infinite-precision Integers, Reals & Modular Integers, IEEE 754 floating point\\ \hline
  Linear Arithmetic & Linear and Non-linear Arithmetic \\ \hline
  Arithmetic or Bit Vectors & Arithmetic \emph{and} Bit Vectors \\ \hline
\end{tabular}
\end{normalsize}
}
\bigskip
\caption{Formal verification vs. real-world development attributes.}
\label{tbl:fm-vs-real-world}
\end{table}

The ACL2 Translator component of Fig.~\ref{RAC:toolchain} provides a
case study in the bridging of Formal Modeling and Real-World
Development concerns, as summarized in
Table~\ref{tbl:fm-vs-real-world}.  The ACL2 translator converts
imperative RAC code to functional ACL2 code.  Loops are translated
into tail-recursive functions, with automatic generation of measure
functions to guarantee admission into the logic of ACL2 (RAC
subsetting rules ensure that loop measures can be automatically
determined).  Structs and arrays are converted into functional ACL2
records.  The combination of modular arithmetic and bit-vector
operations of typical RAC source code is faithfully translated to
functions supported by Russinoff's RTL books.  ACL2 is able to reason
about non-linear arithmetic functions, so the usual concern about
formal reasoning about non-linear arithmetic functions does not apply.  
Finally, the RTL books are quite capable of reasoning about 
a combination of arithmetic and bit-vector operations, which is
a very difficult feat for most automated solvers.

Recently, we have investigated the synthesis of  
Field-Programmable GateArray (FPGA) hardware directly from
high-level architecture models, in collaboration with
colleagues at Kansas State University.  The goal of this work 
is to enable the generation of high-assurance hardware and/or
software from high-level architectural specifications expressed in the
Architecture Analysis and Design Language (AADL) \cite{feiler-AADL},
with proofs of correctness in ACL2.

\section{Rust and RAR}

The Rust Programming Language \cite{Rust2018} is a modern, high-level
programming language designed to combine the code generation
efficiency of C/C++ with drastically improved type safety and memory
management features.  A distinguishing feature of Rust is a non-scalar
object may only have one owner.  For example, one cannot assign a
reference to an object in a local variable, and then pass that
reference to a function.  This restriction is similar to those imposed
on ACL2 single-threaded objects (stobjs) \cite{stobj}, with the
additional complexities of enforcing such ``single-owner''
restrictions in the context of a general-purpose, imperative
programming language.  The Rust runtime performs array
bounds checking, as well as arithmetic overflow checking (the latter
can be disabled by a build environment setting).

In most other ways, Rust is a fairly conventional modern programming
language, with interfaces (called traits), lambdas (termed closures),
and pattern matching, as well as a macro capability.  Also in keeping
with other modern programming language ecosystems, Rust features a
language-specific build and package management sytem, named cargo.

\subsection{Restricted Algorithmic Rust}

As we wish to utilize the RAC toolchain as a backend in our initial
work, the Restricted Algorithmic Rust is semantically equivalent to
RAC.  Thus, we adopt the same semantic restrictions as described in
Russinoff's book.  Additionally, in order to enable translation to RAC,
as well as to ease the transition from C/C++, RAR supports a
commonly used macro that provides a C-like \emph{for} loop in Rust.  
Note that, despite the restrictions, RAR code is proper Rust; it
compiles to binary using the standard Rust compiler.

\begin{figure}
\center{
\begin{tikzpicture}[auto, scale=1.2]
\node[level 6] (Rust) at (-3.2,1.7) [shape=rectangle,draw,align=center]{Rust\\Source};
\node[level 2] (Plexi) at (-0.8,1.7) [shape=rectangle,draw,align=center]{Plexi\\Transpiler};
\node[level 5] (Hdr) at (2,3.2) [shape=rectangle,draw,align=center]{Algorithmic C\\Headers};
\node[level 5] (AlgoC) at (2,1.7) [shape=rectangle,draw,align=center]
{Algorithmic  C\\Source};
\node[level 0] (C++) at (1,0) [shape=rectangle,draw,align=center]{C++\\Compiler};
\node[level 0] (HW) at (3,0) [shape=rectangle,draw,align=center]{Hardware\\Synthesis};
\node[level 2] (Xlate) at (4.7,1.7) [shape=rectangle,draw,align=center]{ACL2\\Translator};
\node[level 4] (Lemmas) at (7,3.2) [shape=rectangle,draw,align=center]{Lemmas};
\node[level 2] (ACL2) at (7,1.7) [shape=rectangle,draw,align=center]{ACL2\\Theorem\\Prover};
\node[level 4] (Proofs) at (7,0.1) [shape=rectangle,draw,align=center]{Proofs};
\draw [->,thick] (Rust) to (Plexi);
\draw [->,thick] (Plexi) to (AlgoC);
\draw [->,thick] (Hdr) to (AlgoC);
\draw [->,thick] (AlgoC) to (C++);
\draw [->,thick] (AlgoC) to (HW);
\draw [->,thick] (AlgoC) to (Xlate);
\draw [->,thick] (Xlate) to (ACL2);
\draw [->,thick] (Lemmas) to (ACL2);
\draw [->,thick] (ACL2) to (Proofs);
\end{tikzpicture}}
\caption{Restricted Algorithmic Rust (RAR) toolchain.}
\label{RAR:toolchain}
\end{figure}
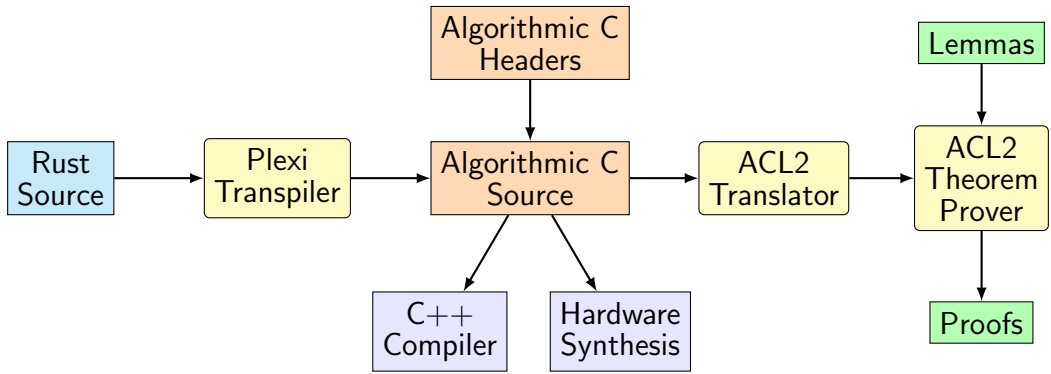

RAR is transpiled to RAC via a source-to-source translator, as
depicted in Fig.~\ref{RAR:toolchain}.  Our transpiler is based on
the \texttt{plex} parser and lexer generator \cite{plex} source code.
We thus call our transpiler \emph{Plexi}, a nickname given to a famous
(and now highly sought-after) line of Marshall guitar amplifiers of
the mid-1960s.  Plexi performs lexical and syntactic transformations
that convert RAR code to RAC code.  This RAC code can then be
compiled using a C/C++ compiler, fed to an HLS-based FPGA compiler,
as well as translated to ACL2 via the RAC ACL2 translator, as illustrated
in Fig.~\ref{RAR:toolchain}.

\section{Example: Array-Based Set}

In this section, we describe an array-based set implementation using
RAR.  This implementation is based on a challenge problem 
formulated in SPARK, which was in turn a sanitization of source code
produced from a Simulink model created during a Collins high-assurance
development \cite{Hardin2009b}.

\subsection{Arrayset Definitions}\label{arraysetDef}

First, we present the basic RAR declaration for the Arrayset.

\begin{verbatim}
const ARR_SZ: uint = 256;

#[derive(Copy, Clone)]
struct Arrayset {
  anext: [usize; ARR_SZ],
  avals: [i64; ARR_SZ],
  free_head: usize,
  used_head: usize,
}
\end{verbatim}

In this declaration, the array \texttt{avals} holds the set elements,
the \texttt{anext} array contains indices indicating the next element in 
either the free or used list, \texttt{free_head} is the index of the first
element of the free list, and \texttt{used_head} is the index of the
first element of the used list.  Note that indices in Rust are
normally declared to be of the \texttt{usize} type.

The ingenious bit about this particular array-backed set
implementation is the use of a single \texttt{anext} 
array to hold both the free list and the used list.  Each element of
the \texttt{anext} array is in one of the two lists, but not both.  The
free list and used list are both terminated by elements in the next
array with the value \texttt{ARR_SZ}.  Any value outside the range
of valid element indices would do for the terminators.  One of the jobs
of the mutators \texttt{aset_init()}, \texttt{aset_add()}, and
\texttt{aset_del()} is to maintain the integrity of the free list and used list
contained within the single \texttt{anext} array, and it is a primary obligation
of the ACL2 proofs to show that this is the case.

\begin{figure}
\begin{center}
\includegraphics{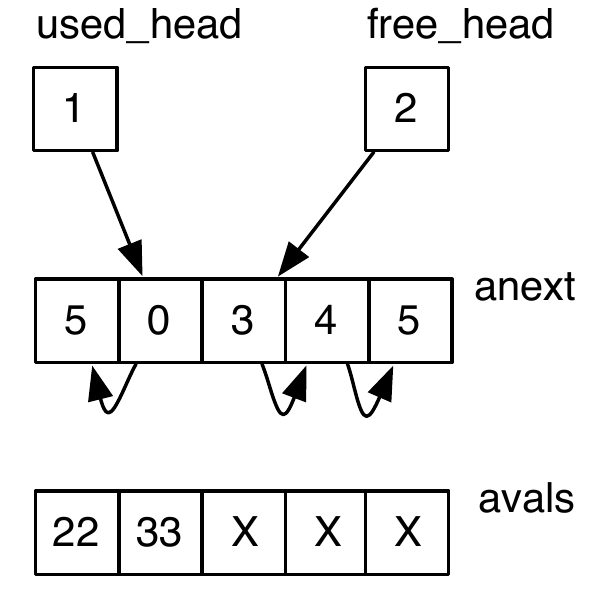}
\end{center}
\caption{An Arrayset data structure with contents \{33, 22\}, size = 5.}
\label{arrayset-pic}
\end{figure}

Fig. \ref{arrayset-pic} shows the state of an Arrayset of size 5 after 
some number of \texttt{aset_add} and \texttt{aset_del} operations
have been performed.  The \texttt{used_head} is 1, which is an index
into the \texttt{anext} array (note that all arrays are zero-based).
The \texttt{free_head} is 2.  If we read the \texttt{anext} array at
index \texttt{used_head}, we get the next element in the used list,
namely 0.  If we then read the \texttt{anext} array at index 0, we get
5, which is the terminator; thus, we have arrived at the end of the
used list.  The corresponding elements in the \texttt{avals} array
are 33 and 22; thus, the Arrayset content is \{33, 22\}. (Note that all
other components of \texttt{avals} have values, but since they are
free elements, those values do not matter.)

On the other hand, if we traverse the \texttt{anext} array
starting at the \texttt{free_head} index, we get 3.  Following the
free list, we read the third element of the \texttt{anext} list, which
is 4.  Reading index 4 of the \texttt{anext} list, we get 5, which
is the terminator.  Thus, we have reached the end of the free list.

\begin{figure}
\begin{center}
\includegraphics{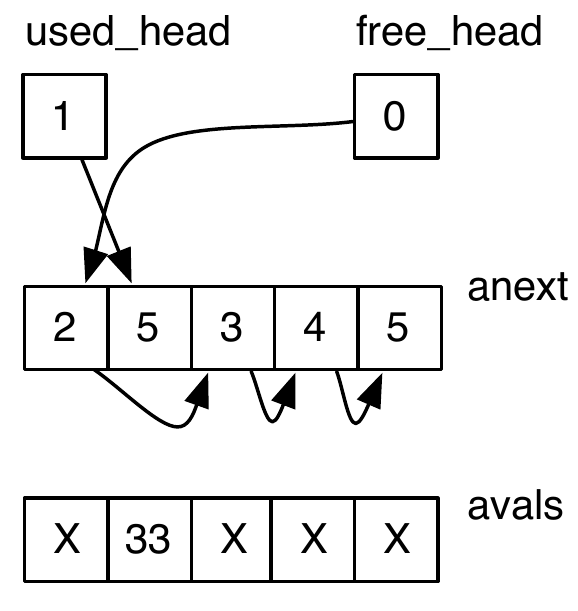}
\end{center}
\caption{The Arrayset after deleting element 22.}
\label{arrayset-del-pic}
\end{figure}

If we then execute \texttt{aset_del(22, aset)}, we obtain
the Arrayset shown in Figure \ref{arrayset-del-pic}.  As one can
readily observe, the used list is shortened by one, and the free list is
lengthened by one, all using elements from the single \texttt{anext} array.

\begin{figure*}
\begin{verbatim}
fn aset_add(val: i64, aset: mut Arrayset) -> Arrayset {
  let curr_index: usize = aset.free_head;

  if (curr_index >= ARR_SZ) {
    return aset;                 // Full
  } else {
    if ((aset.used_head < ARR_SZ) && aset_is_element(val, aset)) {
      return aset;
    } else {

      aset.free_head = aset.anext[aset.free_head];
      aset.avals[curr_index] = val;
      aset.anext[curr_index] = aset.used_head;
      aset.used_head = curr_index;

      return aset;
    }
  }
}
\end{verbatim}
\hrulefill
\caption{aset_add() function in RAR.}
\label{aset-add}
\end{figure*}

The \texttt{aset_add()} function is coded in RAR as shown in
Fig.~\ref{aset-add}.  The operation of \texttt{aset_add()} is as follows.  If the
free head has the value of the terminator, then there is no room, and
no change is made to the Arrayset.  Also, if the element \texttt{val}
to be added is already in the set, then the original Arrayset is
returned.  Otherwise, we adjust the free head to the next element in
the free list, insert the new \texttt{val} at the element index
indicated by the old free head, copy the old free head value to the used 
head position in the next array, and finally set the used head to the old
free head.  Aside from the obvious syntactic differences, this function is a
fairly straightforward translation of the SPARK version of \cite{Hatcliff2011}.

\texttt{aset_del()} proceeds similarly, although it is admittedly
a bit more complicated (see Fig.~\ref{aset-del}).
\texttt{aset_del()}, in turn, depends on the
\texttt{aset_element_prev_from()} function 
(not shown), which searches for an element with a given
value \texttt{val} by traversing the \texttt{anext} array, and
examining values using corresponding indices into the
\texttt{avals} array.  This function actually returns an index
\texttt{previ} such that \texttt{aset.avals[aset.anext[previ]] == val},
as that index is needed for Arrayset bookkeeping.

\begin{figure*}
\begin{verbatim}
fn aset_del(val: i64, aset: mut Arrayset) -> Arrayset {
  let mut curr_index: usize = aset.used_head;
  let mut prev_index: usize;

  if (aset.used_head >= ARR_SZ) {
    return aset;                  // Empty
  } else {

    if (aset.avals[curr_index] == val) {
      aset.used_head = aset.anext[curr_index];
      aset.anext[curr_index] = aset.free_head;
      aset.free_head = curr_index;

      return aset;
    } else {
      prev_index = aset_element_prev_from(aset.used_head, val, aset);

      if (prev_index >= ARR_SZ) {
        return aset;
      } else {

        curr_index = aset.anext[prev_index];

        if (curr_index >= ARR_SZ) {
          return aset;
        } else {
          aset.anext[prev_index] = aset.anext[curr_index];
          aset.anext[curr_index] = aset.free_head;
          aset.free_head = curr_index;

          return aset;
        }
      }
    }
  }
}
\end{verbatim}
\hrulefill
\caption{aset_del function in RAR.}
\label{aset-del}
\end{figure*}

\subsection{Translation to ACL2}

We use \texttt{Plexi} to transpile the RAR source to RAC (not
shown), then use the RAC translator to convert the resulting RAC
source to ACL2.  The translation of \texttt{aset_add()} appears in
Fig.~\ref{aset-add-acl2}.

\begin{figure*}
\begin{verbatim}
(DEFUND ASET_ADD (VAL ASET)
    (LET ((CURR_INDEX (AG 'FREE_HEAD ASET)))
         (IF1 (LOG>= CURR_INDEX (ARR_SZ))
              ASET
              (IF1 (LOGAND1 (LOG< (AG 'USED_HEAD ASET) (ARR_SZ))
                            (ASET_IS_ELEMENT VAL ASET))
                   ASET
                   (LET* ((ASET (AS 'FREE_HEAD
                                    (AG (AG 'FREE_HEAD ASET)
                                        (AG 'ANEXT ASET))
                                    ASET))
                          (ASET (AS 'AVALS
                                    (AS CURR_INDEX VAL (AG 'AVALS ASET))
                                    ASET))
                          (ASET (AS 'ANEXT
                                    (AS CURR_INDEX (AG 'USED_HEAD ASET)
                                        (AG 'ANEXT ASET))
                                    ASET)))
                         (AS 'USED_HEAD CURR_INDEX ASET))))))
\end{verbatim}
\hrulefill
\caption{aset_add() function translated to ACL2 using the RAC tools.}
\label{aset-add-acl2}
\end{figure*}

The first thing to note about Fig.~\ref{aset-add-acl2} is that struct
and array `get' and `set' operations become untyped record operators,
\texttt{AG} and \texttt{AS}, respectively --- these are slight
RAC-specific customizations of the usual ACL2 untyped record operators.
Further, \texttt{IF1} is a RAC-specific  macro, and \texttt{LOG>=}, \texttt{LOGAND1},
and \texttt{LOG<} are all RTL functions.  Thus, much of the proof effort
involved with RAR code is reasoning about untyped records and RTL --- 
although not a lot of RTL-specific knowledge is needed, at least in our
experience.  An additional observation to make here is
that, even though we are two translation steps away from the original
RAR source, the translated function is nonetheless quite readable,
which is a rare thing for machine-generated code.

\subsection{Arrayset Theorems}\label{arraysetThms}

Once we have translated the Arrayset functions into ACL2, we can begin
to prove theorems about the data structure implementation.  We start 
by introducing an important relation between the free head
and the used head that we expect all operations to maintain:

\begin{verbatim}
(defun free-head-used-head-relation (aset)
  (not (= (ag 'free_head aset) (ag 'used_head aset))))
\end{verbatim}

We then define a ``good state'' predicate.  This function 
states that for a good state, the \texttt{aset} input satisfies 
the \texttt{arraysetp} wellformedness predicate; the \texttt{free_head} is not
equal to the \texttt{used_head}, there are no duplicates of
\texttt{val} in the \texttt{avals} array, and that the length of the
free list plus the length of the used list is equal to the total length.

\begin{verbatim}
(defun good-statep (val aset)
  (and (arraysetp aset)
       (free-head-used-head-relation aset)
       (no-dups val aset)
       (= (+ (aset_len aset) (aset_len_free aset)) (arr_sz))))
\end{verbatim}

Given this definition of a good Arrayset state, we can prove
functional correctness theorems for
the Arrayset operations, of the sort stated below:

\begin{verbatim}
(defthm aset_add-works--thm
  (implies
   (and (good-statep val aset)
        (integerp val)
        (< (aset_len aset) (arr_sz)))
   (= (aset_is_element val (aset_add val aset)) 1)))
\end{verbatim}

ACL2 performs the correctness proof for \texttt{aset_add}
automatically.  The correctness proof for the \texttt{aset_del} operation is
considerably more complex, as \texttt{aset_del} makes numerous
modifications to the Arrayset data structure.  Thus, user assistance
is currently required in order to perform the correctness proof.

\section{Comparison to other Automated Verification Tools}\label{comparison}

The Arrayset example has so far been subjected to automated formal
verification by two other approaches: model-checking \cite{Hardin2009b} and symbolic
execution \cite{Hatcliff2011}.  In the former case, the Arrayset
example, as well as its correctness properties, were formulated as
Simulink/Stateflow \cite{Simulink} models (indeed, this was the 
original form), and processed by the Collins Gryphon toolset 
\cite{Gryphon:CACM}.  In the particular instantiation of
the Gryphon toolset in use at the time, the backend model checker
was able to automatically establish correctness properties for the
Arrayset example, but only up to an array size of 3; beyond that,
state space explosion occurred  \cite{Hardin2009b}.

Later, John Hatcliff's team at Kansas State University utilized the
Arrayset SPARK code, automatically produced from the earlier Simulink
model by another part of the Gryphon toolchain \cite{Hardin2009b}, as a challenge
problem for their newly-developed Bakar Kiasan symbolic
execution-based formal analysis tool.  Bakar Kiasan processes
an extended SPARK contract annotation language that includes,
among other enhancements, user-defined correctness predicates 
expressed using SPARK syntax.  Bakar Kiasan was able to prove
functional correctness for the Arrayset example up to an array size of
8 before combinatorics began to overwhelm its symbolic execution
engine \cite{Hatcliff2011}.

Our current work also can be compared and contrasted to a previous
formalization of the array-backed set we developed using ACL2's
single-threaded object capability \cite{arrayset-stobj}.  Whereas the
use of a stobj made for a much more performant ACL2 implementation,
which was useful for validation testing, the stobj-baased
formalization had to be constructed by hand.  Thus, no direct
connection to an imperative implementation, of the sort commonly
written by a non-specialist developer, could be made, in contrast to
the current approach.  Further, reasoning about untyped records is
considerably easier than reasoning about stobj's, at least in our
experience.  We additionally note that in both ACL2
formalizations, all Arrayset operations were written in tail-recursive
style, so this was not a factor in the ease of proof.

\section{Conclusion}\label{Conclusion}

We have developed a prototype toolchain to allow the Rust
programming language to be used as a hardware/software co-design
and co-assurance language for critical systems, standing on the shoulders of 
Russinoff's team at Arm, and all the great work they have
done on Restricted Algorithmic C.  We have demonstrated the ability to
establish the correctness of several practical data structures commonly found
in high-assurance systems (\eg, array-backed singly-linked lists,
doubly-linked lists, stacks, and dequeues) through automated formal verification,
enabled by automated source-to-source translation from Rust to RAC to
ACL2, and we detailed the specification and verification of one such
data structure, an array-backed set.  We have also successfully
applied our toolchain to cryptography and data format filtering
examples typical of the sorts of algorithms that one encounters in
critical systems development.

In the particular case of an array-backed set, we were able to compare
the scalability of several verification techniques, including model-checking,
symbolic execution, and the ACL2 theorem prover.  The model checking
and symbolic execution approaches were able to prove correctness for
small array sizes (up to 3, and up to 8, respectively), but state
space explosion did not allow them to go beyond that.  While the ACL2
proof efforts required a fair amount of human labor in order to
achieve the Arrayset formalization and proofs, now that this effort
has been made, additional proof efforts involving similar data
structures should proceed more quickly.

In future work, we will continue to develop our toolchain, increasing
the number of Rust features that we can support in the RAR subset, as well
as continuing to improve the ACL2 verification libraries in order to
increase the ability to discharge RAR correctness proofs
automatically.  We will also continue to work with our colleagues
at Kansas State University on the direct synthesis and verification of
FPGA hardware from RAR source code.

\section{Acknowledgments}\label{Acknowledgments}

This work was funded by DARPA contract HR00111890001. The
views, opinions and/or findings expressed are those of the authors
and should not be interpreted as representing the official views or
policies of the Department of Defense or the U.S. Government.

Many thanks to Mike Whalen of Amazon for formulating the original
version of the Arrayset example while a Collins employee; to John
Hatcliff of Kansas State University for taking on the
Arrayset example as a challenge problem for his symbolic execution
framework; to Matt Kaufmann at the University of Texas at Austin for
his help on the earlier stobj-based version of the Arrayset code;
to David Russinoff of Arm for answering questions about the RAC toolchain;
and to Robby and Matthew Weis of Kansas State University for their
ongoing work to further our hardware/software co-assurance efforts.
Thanks also go to the anonymous reviewers for their 
insightful comments.

\bibliographystyle{eptcs} 
\bibliography{biblio}

\end{document}